\documentclass[runningheads]{llncs}

\usepackage{graphicx}

\usepackage{times}
\usepackage{soul}
\usepackage{url}
\usepackage[hidelinks]{hyperref}
\usepackage[utf8]{inputenc}
\usepackage[small]{caption}
\usepackage{graphicx}
\usepackage{amsmath}

\usepackage{amsthm}
\usepackage{booktabs}
\usepackage{algorithm}
\usepackage{algorithmic}
\usepackage[toc,page]{appendix}
\usepackage{hyperref}
\usepackage{cleveref}

\newcommand{\mycaption}[2]{\caption[#1]{\textbf{#1.} #2}}

\usepackage{xcolor}
\usepackage[switch]{lineno}

\usepackage{amsmath,amssymb,bbm} 
\DeclareMathOperator*{\argmin}{argmin}
\DeclareMathOperator*{\argmax}{argmax}

\newcommand{\br}{\mathit{br}}

\theoremstyle{definition}
\newtheorem*{definition*}{Definition}

\usepackage{booktabs}
\usepackage{subfig}

\newcommand{\citep}{\cite} 

\usepackage{caption} 
\newtheorem{exmp}{Example}

\title{Evaluating Strategy Exploration in Empirical Game-Theoretic Analysis}

\begin{document}

\author{Yongzhao Wang \and
Qiuri Ma \and
Michael P. Wellman}
\authorrunning{Y. Wang et al.}
%
\institute{University of Michigan, Ann Arbor\\
\email{\{wangyzh, qmaai, wellman\}@umich.edu}}

\maketitle

\begin{abstract}
In empirical game-theoretic analysis (EGTA), game models are extended iteratively through a process of generating new strategies based on learning from experience with prior strategies. 
The \textit{strategy exploration} problem in EGTA is how to direct this process so to construct effective models with minimal iteration.
A variety of approaches have been proposed in the literature, including methods based on classic techniques and novel concepts.
Comparing the performance of these alternatives can be surprisingly subtle, depending sensitively on criteria adopted and measures employed.
We investigate some of the methodological considerations in evaluating strategy exploration, defining key distinctions and identifying a few general principles based on examples and experimental observations.
In particular, we emphasize the fact that empirical games create a space of strategies that should be evaluated as a whole.
Based on this fact, we suggest that the \textit{minimum regret constrained profile} (MRCP) provides a particularly robust basis for evaluating a space of strategies, and propose a local search method for MRCP that outperforms previous approaches.
However, the computation of MRCP is not always feasible especially in large games.
In this scenario, we highlight consistency considerations for comparing across different approaches.
Surprisingly, we find that recent works violate these considerations that are necessary for evaluation, which may result in misleading conclusions on the performance of different approaches.
For proper evaluation, we propose a new evaluation scheme and demonstrate that our scheme can reveal the true learning performance of different approaches compared to previous evaluation methods.

\end{abstract}

\section{Introduction}
Recent years have witnessed dramatic advances in developing game-playing strategies through iterative application of (deep) reinforcement learning (RL)\@.
DeepMind's breakthroughs in Go and other two-player strategy games \citep{silver2017go,silver18short} demonstrated the power of learning through self-play.
In self-play the learner generates an improved strategy assuming its opponent plays the current strategy.
For many games, iterating best-response in this manner will cycle or otherwise fail to converge, which has led to consideration of alternative approaches to generate new strategies. 
For example, DeepMind's milestone achievement in the complex video strategy game StarCraft~II entailed an elaborate population-based search approach \citep{Vinyals19short} informed by game-theoretic concepts.%

Many recent works \citep{Lanctot17,balduzzi2019open,muller2019generalized} have likewise appealed to game-theoretic methods to direct iterative RL for complex games. 
At each iteration, a new strategy is generated for one agent through RL, fixing other agents to play strategies from previous iterations.
A general formulation of this approach is the \textit{Policy Space Response Oracle} (PSRO) algorithm \citep{Lanctot17}.
PSRO can be viewed as a form of \textit{empirical game-theoretic analysis} (EGTA) \cite{Tuyls20,Wellman16putting}, a general name for the study of building and reasoning about game models based on simulation.
In EGTA, game models are induced from simulations run over combinations of a particular set of strategies.
The \textit{strategy exploration} problem in EGTA \cite{Jordan10sw} considers how to extend the considered strategy set, based on the current empirical game model.
For example, one natural approach is to compute a Nash equilibrium (NE) of the current model, and generate a new strategy that optimizes payoff when other agents play that equilibrium.
This approach of iteratively extending strategy sets by best-response to equilibrium was introduced by McMahan \emph{et al.} \cite{mcmahan2003planning} for two-player games and called the \textit{double oracle} (DO) method.

PSRO defines an abstract operation on empirical games, termed 
\textit{meta-strategy solver} (MSS), that extracts an opponent profile from the current empirical game as target for the next best-response calculation. 
In this framework, choosing an MSS determines the strategy exploration method.
For example with NE-calculation as MSS in a two-player game, PSRO reduces to DO\@.
An MSS that simply selects the most recently added strategy corresponds to self-play (SP)\@.
A variety of MSSs have been proposed and assessed in the literature on iterative RL-based approaches to games.
We survey some of these ideas in~\S\ref{sec:related}, as well as alternative approaches to strategy exploration outside the PSRO framework (e.g., not involving RL or varying from best-response).


In practical terms, the proof of a method is whether it produces a superior solution (e.g., a champion Go program).
However, we also seek to understand the relative effectiveness of strategy exploration methods across problem settings, and this remains an open problem for EGTA methodology.
Comparing the performance of alternative methods is subtle because each generates a distinct sequence of strategies, and thus the empirical game model at any point reflects a distinct strategy space.
The relevant comparisons are across different strategy spaces, which may not be faithfully represented by a simple summary such as an interim solution.
This fact has tended to be neglected in studies proposing and evaluating new ideas on strategy exploration \citep{Lanctot17,muller2019generalized}, and as we demonstrate below, this can lead to misleading conclusions on the performance of different approaches.

The present study illuminates several methodological considerations for strategy exploration.
First, we identify a key distinction between PSRO and other learning dynamics that empirical games create a space of strategies that should be evaluated as a whole. Then we seek a theoretically justifiable evaluation metric for empirical games, and suggest the proposal by Jordan \emph{et al.} \cite{Jordan10sw} that the regret of the \textit{minimum-regret constrained-profile} (MRCP) can serve this purpose. 
We show the effectiveness and advantages of using MRCP as a metric through examples. To find MRCP more accurately, we propose a variant of amoeba method \citep{nelder1965simplex} that outperforms previous approaches in matrix games. 
Calculation of MRCP is not always computationally feasible, so we identify some desiderata for alternative evaluation metrics.
Specifically we argue for the importance of evaluating the whole space of strategies in an empirical game, and highlight some consistency considerations for comparing across different MSSs. We point out the MSS used for evaluation is not necessarily the same as the MSS in strategy exploration and define \textit{solver-based regret} for evaluation purposes. 
Based on these considerations, we propose a new evaluation solver selection scheme for EGTA, which leads to a sensible comparison across MSSs.
We demonstrate the significance of our considerations and approaches in both synthetic and real-world games. 

Finally, we consider the problem of regret-based evaluation in situations where calculating exact best response is infeasible thus accurate regret is not available. 
One alternative that is widely applied is using generated strategies for evaluation purpose where regret calculation for different MSSs only considers deviations within the generated strategies. 
We test this approach from a game-theoretic perspective and find that high-regret profiles in the true game may exhibit low regret in the combined game, thus casting doubt on the accuracy of this approach.

Contributions of this study include: 
\begin{enumerate}
    \item Recognition that empirical games create a space of strategies that should be evaluated as a whole. To serve this purpose, we suggest MRCP as evaluation metric and present evidence that MRCP provides a particularly robust basis for evaluation. 
    \item The notion of \textit{solver-based regret} for evaluation, with focus on consistency considerations for comparing MSSs. We demonstrate the potential for misleading results when consistency is violated as in the prior literature. 
    \item A new evaluation solver selection scheme which leads to a sensible comparison across MSSs.
    \item A variant of the amoeba method that outperforms previous approaches in matrix games, plus some insight on MRCP approximation in games wherein regret calculation is restricted; 
    \item An assessment of approaches to evaluate strategy exploration when calculating exact best response is infeasible.
\end{enumerate}

\section{Related Work on Strategy Exploration}
\label{sec:related}

In the first instance of automated strategy generation in EGTA, Phelps \emph{et al.} \cite{Phelps06} employed genetic search over a parametric strategy space, optimizing performance against an equilibrium of the empirical game.
Schvartzman \& Wellman \cite{Schvartzman09} combined RL with EGTA in an analogous manner.
Questioning whether best response to equilibrium is an ideal way to add strategies, these same authors framed and investigated the general problem of \textit{strategy exploration} in EGTA \citep{Schvartzman09a}.
They identified situations where adding a best response to equilibrium would perform poorly, and proposed some alternative approaches.
Jordan \emph{et al.} \cite{Jordan10sw} extended this line of work by proposing exploration of strategies that maximize the gain to deviating from a rational closure of the empirical game.

Investigation of strategy exploration was furthered significantly by introduction of the PSRO framework \citep{Lanctot17}.
PSRO entails adding strategies that are best responses to \textit{some} designated other-agent profile, where that profile is determined by the \textit{meta-strategy solver} (MSS) applied to the current empirical game.
The prior EGTA approaches cited above effectively employed NE as MSS as in the DO algorithm \citep{mcmahan2003planning}.
Lanctot \emph{et al.} \cite{Lanctot17} argued that with DO the new strategy may overfit to the current equilibrium, and accordingly proposed and evaluated several alternative MSSs, demonstrating their advantages in particular games.
For example, their \emph{projected replicator dynamics} (PRD) employs an RD search for equilibrium \citep{taylor1978evolutionary,smith1973logic}, but truncates the replicator updates to ensure a lower bound on probability of playing each pure strategy.
Any solution concept for games could in principle be employed as MSS, as for example the adoption by Muller \emph{et al.} \cite{muller2019generalized} of a recently proposed evolutionary-based concept, $\alpha$-rank \citep{omidshafiei2019alpha}, within the PSRO framework.

The MSS abstraction also connects strategy exploration to iterative game-solving methods in general, whether or not based on EGTA\@.
Using a uniform distribution over current strategies as MSS essentially reproduces the classic \textit{fictitious play} (FP) algorithm \citep{brown1951iterative}, and as noted above, an MSS that just selects the most recent strategy equates to self-play (SP).
Note that these two MSS instances do not really make substantive use of the empirical game, as they derive from the strategy sets alone.

Wang \emph{et al.} \cite{wang19sywsjf} illustrated the possibility of combining MSSs, employing a mixture of NE and uniform which essentially averages DO and FP\@.
Motivated by the same aversion to overfitting the current equilibrium, Wright \emph{et al.} \cite{Wright19} proposed an approach that starts with DO, but then fine-tunes the generated response by further training against a mix of previously encountered strategies.

In the literature, a profile's fitness as solution candidate is measured by its regret in the true game. 
Jordan \emph{et al.} \cite{Jordan10sw} defined \textit{MRCP} (\textit{minimum-regret constrained-profile}) as the profile in the empirical game with minimal regret relative to the full game. 
Regret of the MRCP provides a measure of accuracy of an empirical game, but we may also wish to consider the coverage of a strategy set in terms of diversity.
Balduzzi \emph{et al.} \cite{balduzzi2019open} introduced the term \textit{Gamescape} to refer to the scope of joint strategies covered by the exploration process to a given point. 
They employed this concept to characterize the effective diversity of an empirical game state, and proposed a new MSS called \textit{rectified Nash} designed to increase diversity of the Gamescape. 
Finally, we take note of a couple of recent works that characterize Gamescapes in terms of topological features. 
Omidshafiei \emph{et al.} \cite{omidshafiei2020navigating} proposed using spectral analysis of the $\alpha$-rank best response graph, and Czarnecki \emph{et al.}  \cite{czarnecki2020real} visualize the strategic topography of real-world games as a spinning top wherein layers are transitive and strategies within a layer are cyclic.

\section{Preliminaries}
A normal-form game $\mathcal{G}=(N,(S_i),(u_i))$ consists of a finite
set of players $N$ indexed by $i$; a non-empty set of strategies
$S_i$ for player $i\in N$; and a utility function $u_i: \prod_{j \in N}S_j \rightarrow \mathbb{R}$ for player $i\in N$, where $\prod$ is the Cartesian product. 

A mixed strategy $\sigma_i$ is a probability distribution over strategies in $S_i$, with $\sigma_i(s_i)$ denoting the probability player~$i$ plays strategy $s_i$. 
We adopt conventional notation for the other-agent profile: $\sigma_{-i}=\prod_{j \ne i}\sigma_j$. 
Let $\Delta(\cdot)$ represent the probability simplex over a set. The mixed strategy space for player~$i$ is given by $\Delta(S_i)$. Similarly, $\Delta(S)=\prod_{i \in N}\Delta(S_i)$ is the mixed profile space.

Player~$i$'s \textit{best response} to profile $\sigma$ is any strategy yielding maximum payoff for~$i$, holding the other players’ strategies constant:
\begin{displaymath}
\br_i(\sigma_{-i}) = \argmax_{\sigma_i'\in \Delta(S_i)} u_i(\sigma_i', \sigma_{-i}).
\end{displaymath}{}
Let $\br(\sigma)=\prod_{i \in N}\br_i(\sigma_{-i})$ be the overall best-response correspondence for a profile~$\sigma$. 
A Nash equilibrium (NE) is a profile $\sigma^*$ such that $\sigma^*\in \br(\sigma^*)$.

Player~$i$'s \textit{regret} in profile $\sigma$ in game $\mathcal{G}$ is given by 
\begin{displaymath}
    \rho^{\mathcal{G}}_i(\sigma) = \max_{s_i'\in S_i}u_i(s_i', \sigma_{-i})-u_i(\sigma_i, \sigma_{-i}).
\end{displaymath}
Regret captures the maximum player~$i$ can gain in expectation by unilaterally deviating from its mixed strategy in $\sigma$ to an alternative strategy in $S_i$\@. 
An NE strategy profile has zero regret for each player. 
A profile is said to be an $\epsilon$-Nash equilibrium ($\epsilon$-NE) if no player can gain more than $\epsilon$ by unilateral deviation. 
The regret of a strategy profile $\sigma$ is defined as the sum over player regrets:%
\footnote{Some treatments employ max instead of sum for this; both are relevant measures of distance from equilibrium.}
\begin{displaymath}
    \rho^{\mathcal{G}}(\sigma) = \sum_{i\in N} \rho^{\mathcal{G}}_i(\sigma).
\end{displaymath}

An \textit{empirical game} $\mathcal{G}_{S\downarrow X}$ is an approximation of the true game $\mathcal{G}$, in which players choose from restricted strategy sets $X_i\subseteq S_i$, and payoffs are estimated through simulation. 
That is, $\mathcal{G}_{S\downarrow X}=(N,(X_i),(\hat{u}_i))$, where $\hat{u}$ is a projection of $u$ onto the strategy space $X$\@.%
\footnote{Because payoffs are estimated through simulation, $\hat{u}$ is also subject to sampling error.
This presents additional statistical issues \citep{Tuyls20,Vorobeychik10,Wiedenbeck14}; here we ignore those and focus on the issues that arise from strategy set restriction.
} 
We use the notation $\rho^{\mathcal{G}_{S\downarrow X}}$ to make clear when we are referring to regret with respect to an empirical game as opposed to the full game.

A meta-strategy solver (MSS), denoted by $h\in H$, is a function mapping from an empirical game to a strategy profile $\sigma$ within the empirical game. 
Examples of MSS (introduced in~\S\ref{sec:related}) include NE, PRD, uniform, etc. 
PSRO employing a given MSS may have an established name (e.g., PSRO with NE is DO, with uniform is FP); otherwise we may simply refer to the overall algorithm by the MSS label (e.g., PRD may denote the MSS or PSRO with this MSS).

\section{Evaluating Strategy Exploration}

The purpose of evaluating strategy exploration is to understand the relative effectiveness of different exploration methods (e.g., MSSs) across different problem settings.
We achieve this purpose through analyzing the intermediate empirical game models they generate during exploration.

\subsection{Evaluating an Empirical Game Model} \label{sec:MRCP}
From the perspective of strategy exploration, the key feature of an empirical game model is what strategies it incorporates.%
\footnote{The accuracy of the estimated payoff functions over these strategies is also relevant, but mainly orthogonal to exploration and outside the scope considered here.}
In EGTA, the restricted strategy set $X$ is typically a small slice of the set of all strategies~$S$, so the question is how well $X$ covers the strategically relevant space.
There may be several ways to interpret ``strategically relevant'', but one natural criterion is whether the empirical game $\mathcal{G}_{S\downarrow X}$ covers solutions or approximate solutions to the true game~$\mathcal{G}$\@.

The profile in the empirical game closest to being a solution of the full game is the MRCP, as described above.
Formally, $\bar{\sigma}$ is an MRCP of $\mathcal{G}_{S\downarrow X}$ iff: 
\begin{equation}
    \bar{\sigma} = \argmin_{\sigma\in \Delta(X)} \sum_{i\in N} \rho^{\mathcal{G}}_i(\sigma)
\end{equation}\label{eq: MRCP def}
The regret of MRCP thus provides a natural measure of how well $X$ covers the strategically relevant space.
In prior literature, MRCP was studied in games with fixed strategy sets rather than a setting where strategy sets are iteratively built. 
We extend the study of its properties to our strategy exploration setting.
We first highlight that the regret of MRCP decreases monotonically as the empirical game model is being extended, since adding strategies can only increase the scope of minimization.
Moreover, MRCP tracks convergence in that the regret of MRCP reaches zero exactly when an NE of $\mathcal{G}$ is contained in the empirical game, that is, $X$ covers the support of the NE\@.
We claim both properties of MRCP are important and desirable for evaluation purposes.

Unfortunately, computing MRCP as a means for evaluating strategy exploration can be computationally challenging.
Calculating regret of a profile, the quantity we are minimizing, generally requires a best-response oracle for the full game, which itself can be quite computationally expensive (which is why we often find RL the best available method). 
And even given an effective way to calculate regret, the search for MRCP is a non-convex optimization problem over the profile space of the empirical game.

\subsection{Solver-Based Regret}

Given the general difficulty of computing MRCP, studies often employ some other method to select a profile from the empirical game to evaluate. 
Any such method can be viewed as a meta-strategy solver, and so we use the term \textit{solver-based regret} to denote regret in the true game of a strategy profile selected by an MSS from the empirical game.
In symbols, the solver-based regret using a particular MSS is given by $\rho^{\mathcal{G}}(\mathit{MSS}(\mathcal{G}_{S\downarrow X}))$.
By definition, MRCP is the MSS that minimizes solver-based regret.

An MSS that is commonly employed for solver-based regret is NE\@.
NE-based regret measures the stability in the true game of a profile that is perfectly stable in the empirical game.
Whereas any MSS is eligible to play the role of solver, not all are well-suited for evaluating strategy exploration.
For example, SP simply selects the last strategy added, and is completely oblivious to the rest of the strategy set $X$\@.
This clearly fails to measure how well $X$ as a whole captures the strategically relevant part of $S$, which is the main requirement of an evaluation measure as described above.

\subsection{Solver Consistency for Evaluation}

Our framework as described to this point employs MSSs in two distinct ways: to direct a strategy exploration process, and to evaluate intermediate results in strategy exploration.
It may seem natural to evaluate exploration that employs MSS $M$ in terms of solver-based regret with $M$ as solver.
Indeed, much prior work in PSRO exploration has done exactly this
\citep{Lanctot17,Li21w,muller2019generalized}.%
\footnote{One of these is a recent paper from our own group \cite{Li21w}.
Although that work is not focused on strategy exploration, it does present some plots (Figs.\ 2 and~3) with multiple curves using different MSSs for evaluating regret.
For others' works, we verified this by examining the published code and through our own efforts to reproduce the results in these papers. 
Specifically, we found the code published as part of OpenSpiel~\citep{lanctot2019openspiel} evaluates progress in exploration by regret of the MSS employed for exploration. 
We also reproduced the learning performance of PSRO with different MSSs and inferred that the MSS used for evaluation is the same as the one for strategy exploration, which is often apparent by examination of regret curves. 
For example, the NE-based regret curve of fictitious play oscillates dramatically while its uniform-based regret curve is much more smooth. 
So it is easy to identify which MSS was used for evaluation.}

As we demonstrate below, however, evaluating alternative MSSs $M$ and $M'$ for exploration using their respective MSSs as solvers can produce misleading comparisons, which is caused by neglecting the empirical game should be evaluated as a whole.
Instead, we argue, one should apply the same solver-based regret measure to evaluate results under $M$ and $M'$\@. 
In other words, the MSS employed in solver-based regret should be fixed and independent of the MSSs employed for exploration.
We term this the \textit{consistency} criterion.

To illustrate the necessity of solver consistency, we offer two examples to demonstrate how a violation of our consistency criterion could lead to a misleading conclusion.  

\begin{exmp} \label{exp: matrix game}
Consider the symmetric zero-sum matrix game of Table~\ref{tab: Matrix game for NashConv}. Starting from the first strategy of each player, we perform PSRO with uniform and NE as MSSs, respectively. 
The first few iterations of PSRO are presented in Table~\ref{tab: PSRO with DO and FP}. 
Due to symmetry, the two players' strategy sets and MSS-proposed strategies are identical. 

Fig.~\ref{fig:Difference between 2 NashConv Synthetic Game} presents regret curves for both MSSs using NE-based regret, as well as the uniform-based regret curve for FP\@.
If we violate the consistency criterion and compare uniform-based regret of FP with the NE-based regret of DO (i.e., green versus blue curves in Fig.~\ref{fig:Difference between 2 NashConv Synthetic Game}), we would conclude FP converges faster than DO in the first two iterations. 
However, FP cannot actually be better at strategy exploration, as the strategies introduced, $a^1$ and $a^3$, are identical under two MSSs. 
Moreover, at the third iteration, FP fails to add any new strategy, and so the improvement shown is not really due to the exploration process.

Comparing the two MSSs under NE-based regret, (i.e., green versus orange regret curves), we see that where FP and DO generate identical empirical games their evaluations coincide. 
Thus, following the rule of consistency avoids the misleading conclusion.

\begin{table}[!ht]
    \centering
    \begin{tabular}{ |c|c|c|c| } 
 \hline
  & $a_2^1$ & $a_2^2$ & $a_2^3$  \\ 
 \hline
 $a_1^1$ & (0, 0) & (-0.1, 0.1) & (-3, 3) \\ 
 \hline
 $a_1^2$ & (0.1, -0.1) & (0, 0) & (2, -2) \\ 
 \hline
 $a_1^3$ & (3, -3) & (-2, 2) & (0, 0) \\ 
 \hline
\end{tabular}
\caption{A symmetric zero-sum game (Example~\ref{exp: matrix game}).}
\label{tab: Matrix game for NashConv}
\end{table}{}

\begin{table}[!ht]
    \centering
    \begin{tabular}{ccc}
        \toprule
        \textbf{Iter\#} & \textbf{Strategy Sets} & \textbf{DO proposed strategy} \\
        \midrule
        1 & ($a_1^1$), ($a_2^1$) & $(1),(1)$ \\
        2 & ($a_1^1,a_1^3$), ($a_2^1,a_2^3$) & $(0,1),(0,1)$\\
        3 & ($a_1^1,a_1^2,a_1^3$), ($a_2^1,a_2^2,a_2^3$) & $(0,1,0),(0,1,0)$\\
        \bottomrule
         & & \\
        \toprule
        \textbf{Iter\#} & \textbf{Strategy Sets} & \textbf{FP proposed strategy} \\
        \midrule
        1 & ($a_1^1$),($a_2^1$) & $(1),(1)$ \\
        2 & ($a_1^1,a_1^3$),($a_2^1,a_2^3$) & $(\frac{1}{2},\frac{1}{2}),(\frac{1}{2},\frac{1}{2})$\\
        3 & ($a_1^1,a_1^3$),($a_2^1,a_2^3$) & $(\frac{1}{3},\frac{2}{3}),(\frac{1}{3},\frac{2}{3})$\\
        4 & ($a_1^1,a_1^2,a_1^3$),($a_2^1,a_2^2,a_2^3$) & $(\frac{1}{4},\frac{1}{4},\frac{1}{2}),(\frac{1}{4},\frac{1}{4},\frac{1}{2})$\\
        5 & ($a_1^1,a_1^2,a_1^3$),($a_2^1,a_2^2,a_2^3$) & $(\frac{1}{5},\frac{2}{5},\frac{2}{5}),(\frac{1}{5},\frac{2}{5},\frac{2}{5})$\\
        \bottomrule
    \end{tabular}
\mycaption{PSRO process for DO (top) and Fictitious Play (bottom)}{Items are arranged according to the index of players.}
\label{tab: PSRO with DO and FP}
\end{table}{}

\begin{figure*}[!htbp]
\centering
    \subfloat[$3\times 3$ game (Example~\ref{exp: matrix game}).]{\includegraphics[width=0.49\textwidth]{  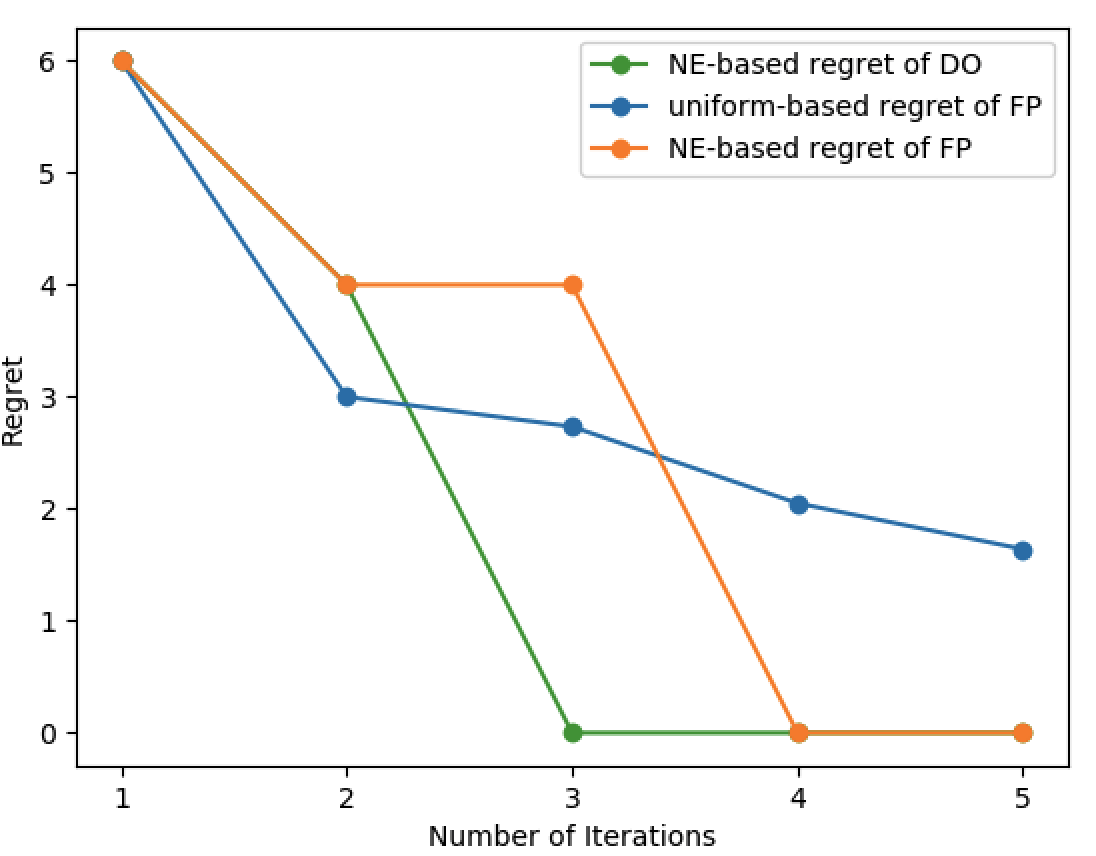}\label{fig:Difference between 2 NashConv Synthetic Game}}\
    \subfloat[$100\times 100$ game (Example~\ref{ex:100game}).]{\includegraphics[height=4.74cm,width=0.49\textwidth]{ 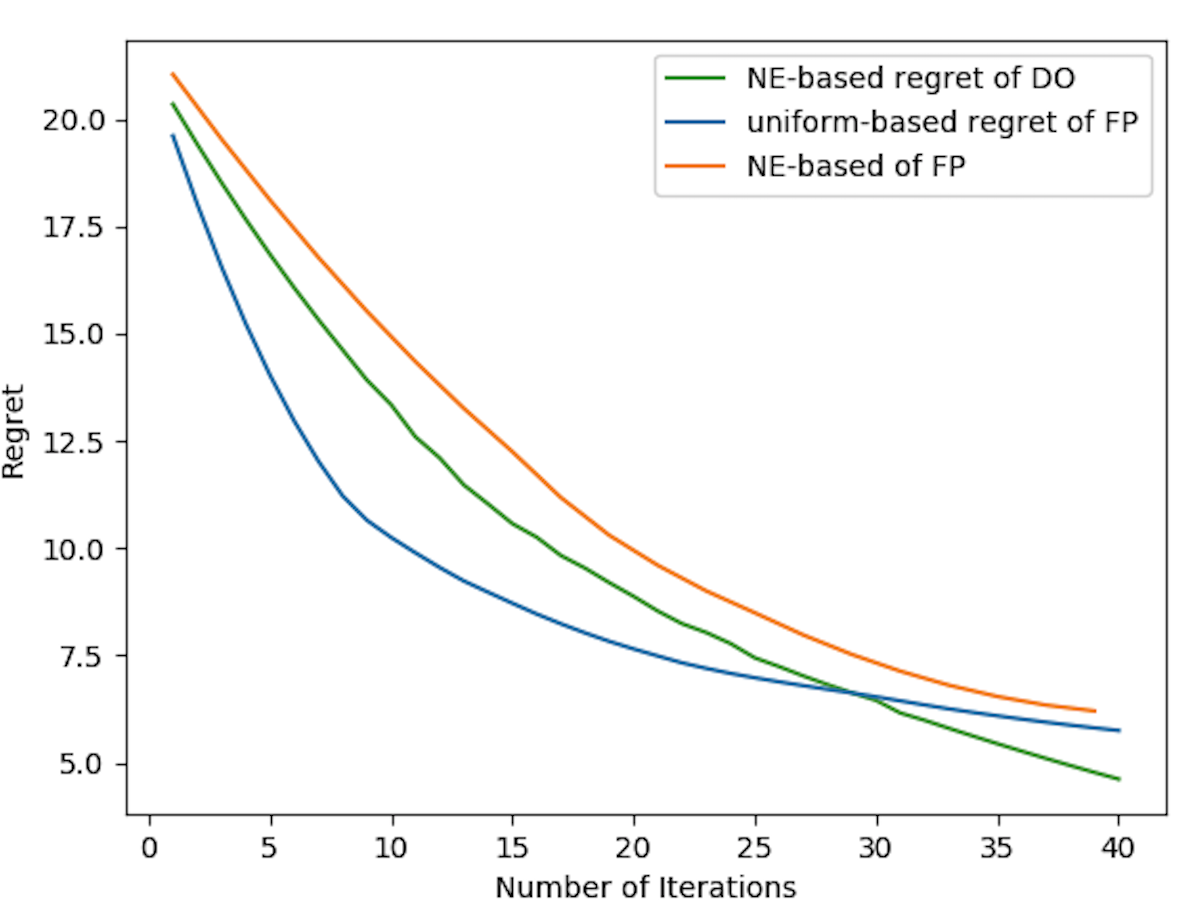}\label{fig:Difference between 2 NashConv Matrix Game}}
\caption{Regret curves evaluating NE and uniform as MSSs strategy exploration, under different solvers.}
\end{figure*}

\end{exmp}

\begin{exmp}\label{ex:100game}
We further verify our observations in a synthetic zero-sum game with 100 strategies per player. 
Resulting regret curves averaged over 10 random starts are shown in Fig.~\ref{fig:Difference between 2 NashConv Matrix Game}. 
As for the previous example, comparing uniform-based regret of FP against NE-based regret of DO would lead us astray, in this case indicating that FP performs better than DO over the first thirty iterations. 
However, the conclusion reached by breaking our consistency criterion is in fact invalid. 
Specifically, before 30 iterations, the uniform-based regret curve indicates the uniform strategy profile in the empirical game of FP is more stable, i.e. with lower regret, than NE in the empirical game of DO. 
Nevertheless, it is very much likely there exists another profile in the empirical game of DO that has far lower regret in the true game, which we demonstrate in the next section. 

This example demonstrates mixed use of evaluation metrics may result in improper comparison among the performance of MSSs.
Indeed, we have found that this phenomenon is quite common in prior work, leading in particular to misleading evaluations of FP as a strategy exploration approach.
In formulating the general consistency criterion, we emphasize that improper comparisons could be made with any two MSSs; the issue is not limited to FP or any specific MSSs employed in these examples.

\end{exmp}

\subsection{Evaluation Solver Selection}

We further examine the consistency criterion in simplified poker games, specifically 2-player Kuhn's poker and Leduc poker.
These poker games have been widely employed in prior work within the PSRO framework, facilitating comparison of experimental results.
Specifically, we evaluate FP, PRD, and NE as MSSs.
Moreover, to select an effective solver to implement the consistency criterion, we propose a new evaluation solver selection scheme, which effectively reveals the authentic performance of MSSs.

\begin{figure*}[!ht]
\centering
    \subfloat[Fictitious Play in 2-player Leduc]{\includegraphics[height=3.74cm, width=0.49\linewidth]{ 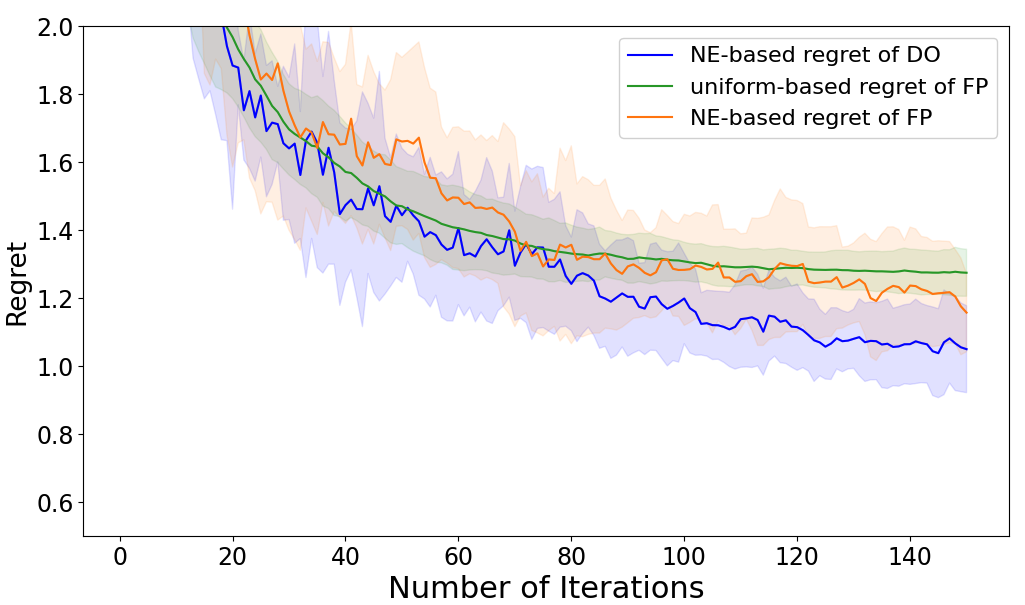}\label{fig:PSRO with fic Leduc}}\
    \subfloat[Fictitious Play in 2-player Kuhn]{\includegraphics[height=3.71cm,width=0.49\linewidth]{ 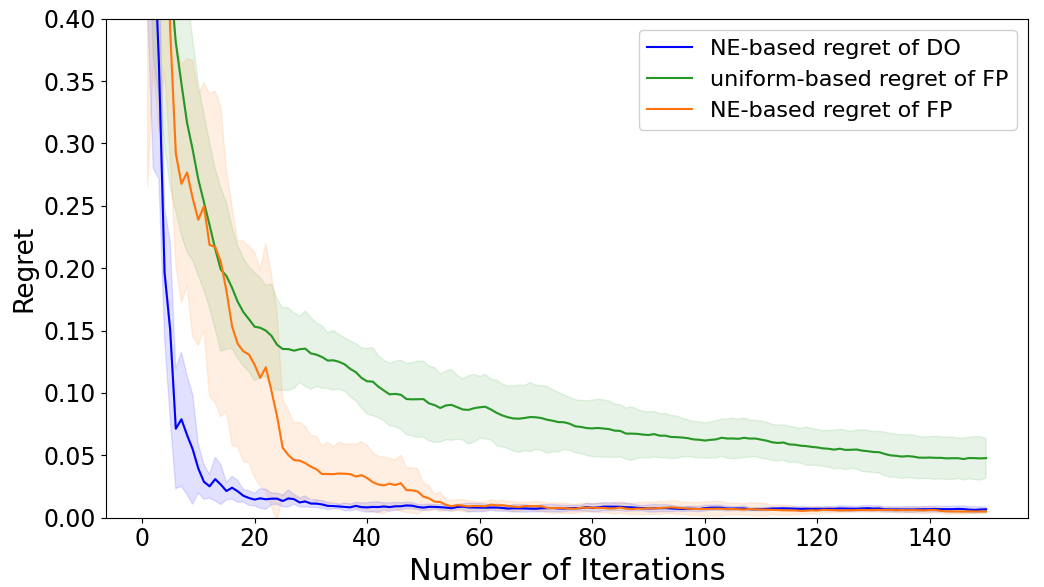}\label{fig:PSRO with fic Kuhn}}\\
    \subfloat[PRD in 2-player Leduc]{\includegraphics[width=0.49\linewidth]{ 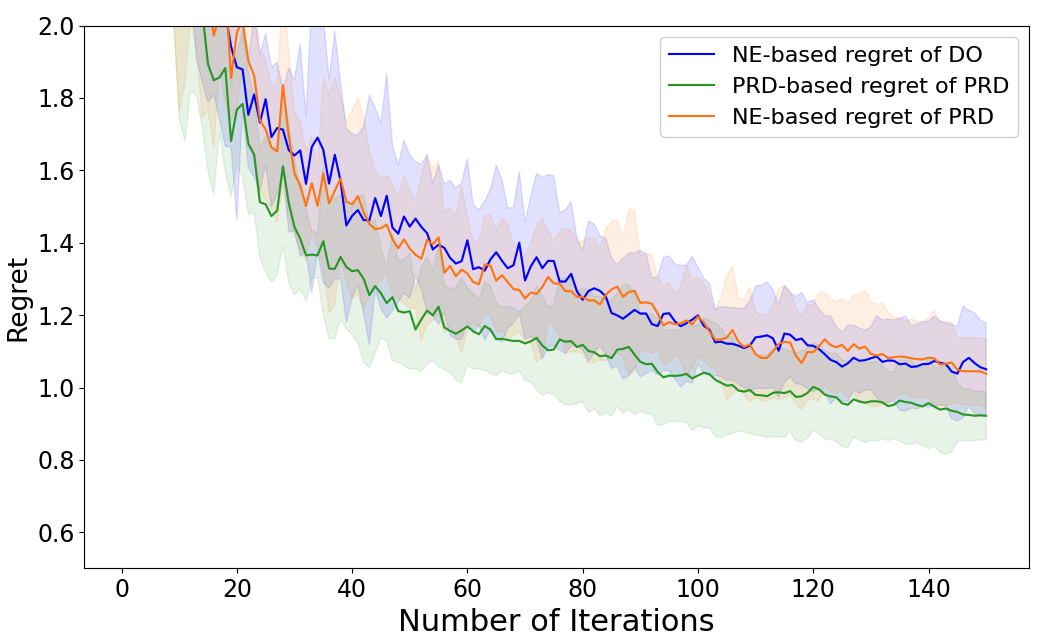}\label{fig:PSRO with prd Leduc}}\
    \subfloat[PRD in 2-player Kuhn]{\includegraphics[height=3.68cm,width=0.49\linewidth]{ 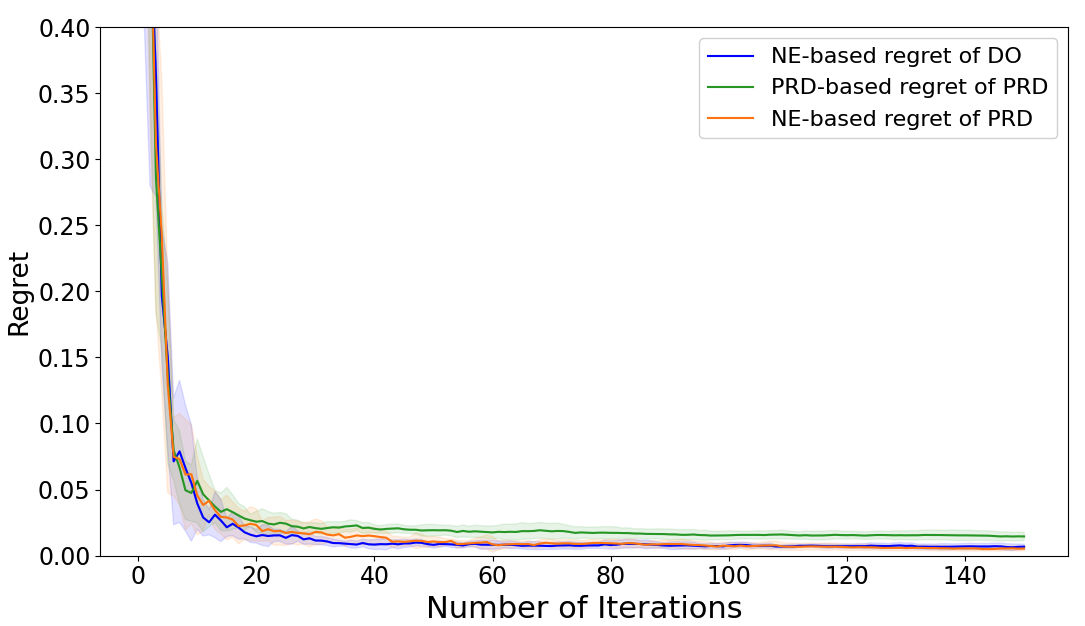}\label{fig:PSRO with prd Kuhn}}\\
    \subfloat[PRD strategies in DO run ]{\includegraphics[width=0.49\linewidth]{ 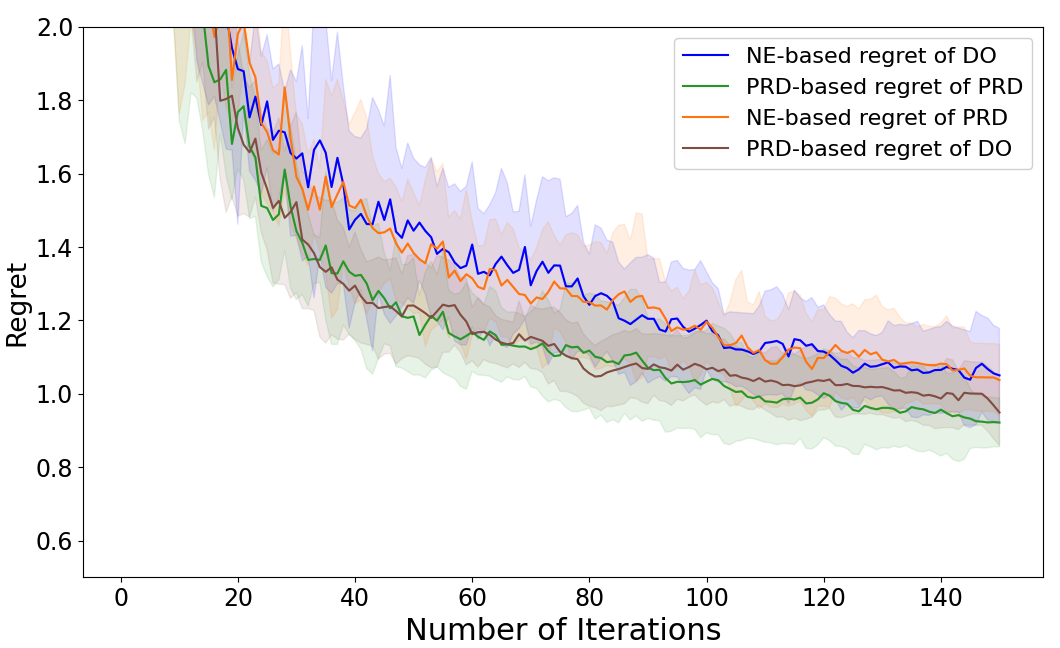}\label{fig:PSRO with prd Leduc 4 curves}}\
    \subfloat[Regret curves by Combined Games]{\includegraphics[height=3.68cm, width=0.49\linewidth]{ 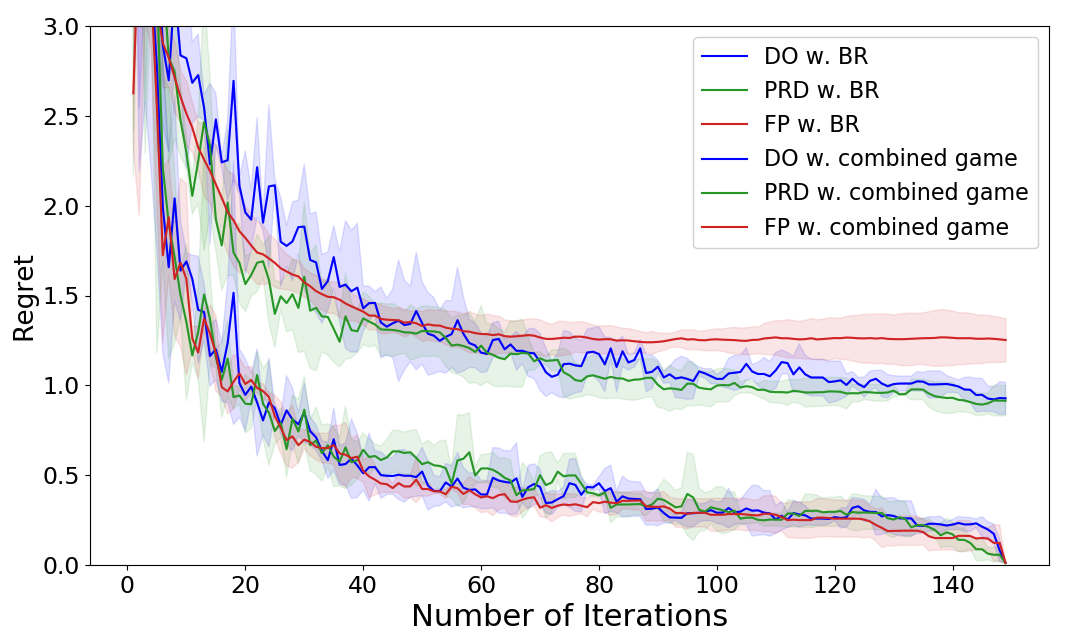}\label{fig:Combined Game regrets}}
\caption{Experimental regret curves for poker games.}\label{fig: Poker Results}
\end{figure*}

\subsubsection{Solver Consistency with FP}
For Leduc poker, Fig.~\ref{fig:PSRO with fic Leduc} indicates DO performs better than FP, regardless of whether the latter is evaluated by NE-based or uniform-based regret and there is little difference between the two regret curves. 
In Kuhn's poker (Fig.~\ref{fig:PSRO with fic Kuhn}), however, FP shows much faster convergence under NE-based rather than uniform-based regret after 20 iterations or so.
Indeed, the uniform-based regret is far from zero even at a hundred iterations.
We saw in the examples above that uniform-based evaluation of FP may misleadingly show smooth improvement where there is none.
Here we see that it can also misleadingly leave the impression of slow progress even when FP has actually introduced the key strategies needed for accurate solution. 
Either way, solver consistency helps to ensure a more meaningful comparison.

\subsubsection{Solver Consistency with PRD}
We reproduce the PSRO experiments with PRD in Leduc poker in Fig.~\ref{fig:PSRO with prd Leduc}. 
We first note that following the rule of consistency, there is little performance gap between PRD and DO (i.e., the blue and orange curves). 
If we violate consistency and compare PRD-based regret of PRD against NE-based regret of DO (green versus blue curve), however, we would be prone to conclude incorrectly that PRD clearly outperforms DO\@.
This again illustrates that using different evaluation solvers could give distinct conclusions on the performance.

The above examples have shown the importance of the consistency criterion with some evaluation solvers. 
In fact, not all MSSs are equally suited for evaluation with the consistency criterion.
To select a meaningful evaluation solver, we propose a scheme, which effectively reveals the true performance of different MSSs.

\subsubsection{An Evaluation Solver Selection Scheme}
Recall that the regret of MRCP of an empirical game provides a natural measure of how well the empirical game covers the strategically relevant space and that MRCP is the MSS minimizing solver-based regret by definition. 
Therefore, one key step to determine which solver to use is to identify MRCP or approximate MRCP.
Fulfilling this spirit, we propose a heuristic evaluation solver selection scheme, which chooses the solver with lowest-regret curve among running solvers, thus approximating MRCP. We demonstrate the significance of our scheme for evaluating different MSSs by checking the previous PRD example.

In the example, if we merely follow the solver consistency with NE-based regret, we may not distinguish the performance difference between PRD and DO, i.e., blue versus orange regret curves with NE-based regret.
This is because NE in the empirical game may have very high regret with respect to the true game, thus far from MRCP and not qualified for being an evaluation solver.
This phenomenon supports our claim that not all MSSs are equally suited for evaluation with the consistency criterion.

To reveal the true relationship between PRD and DO, we apply our solver selection scheme.
With the scheme, we first observe that PRD-based regret of PRD is lower than NE-based regret of DO, i.e., green versus blue regret curves, thus we select PRD as the solver for evaluation and use the PRD-based regret accordingly.
Following the solver consistency, we add the PRD-based regret of DO to Fig.~\ref{fig:PSRO with prd Leduc}, as shown in Fig.~\ref{fig:PSRO with prd Leduc 4 curves}.
Note that PRD as an evaluation solver successfully identifies the profiles with lower regret in the empirical game of DO, i.e., PRD-based regret of DO is lower than NE-based regret of DO, which matches our design purpose of searching profiles close to MRCP.

By comparing the PRD-based regret curves of DO and PRD, we observe that despite they share the similar convergence rate at the initial learning stage, PRD becomes better as learning proceeds in terms of lower regret and variance, which is the authentic relationship between PRD and DO.
Our results imply that the performance gap between PRD and DO is not as large as that without following solver consistency, which again emphasizes that the violation of solver consistency may artificially enlarge performance gap.

Based on the observations, we claim that there could exist more stable strategy profiles than the ones proposed by the solver in an empirical game, as the PRD strategy profiles being more stable than NE profiles in DO. 
This highlights the importance of focusing on evaluating the strategy population as a whole rather than a single profile in the empirical game. 
For Kuhn's poker shown in Fig.~\ref{fig:PSRO with prd Kuhn}, the difference between regret curves are minimal.  

Finally, for situations where the regret curves interleave, fixing an evaluation solver for the whole learning procedure may not be proper. 
In this scenario, our scheme could be implemented by alternating evaluation solvers for different learning phases, insisting on the motivation of searching towards MRCP.

\section{Performance of MRCP and Calculation Refinement} \label{sec: MRCP calculation}

\subsection{Evaluation Performance of MRCP}
\begin{figure}[!ht]
\centering
\includegraphics[width=0.6\columnwidth]{ 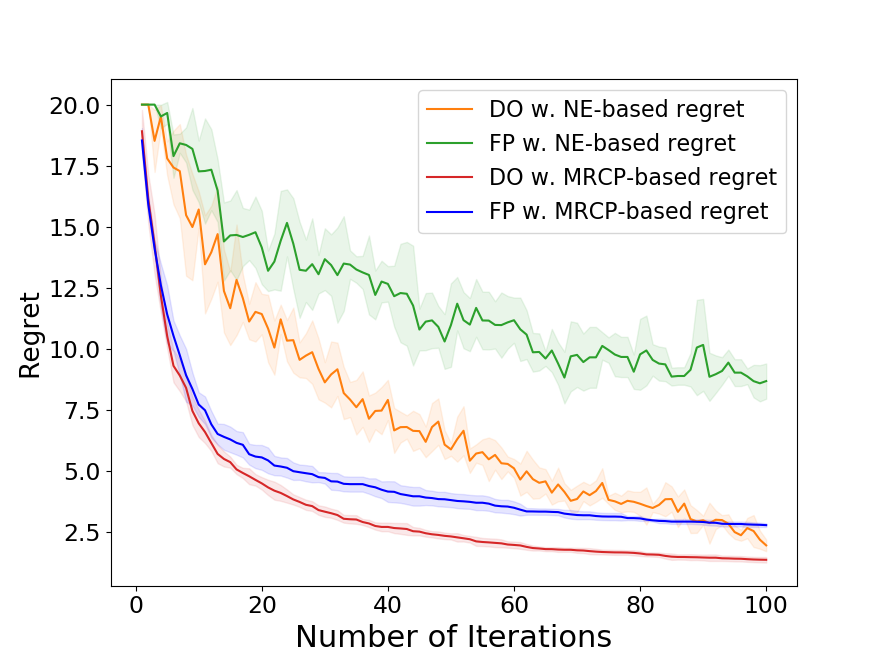}
\caption{MRCP-based Regret vs NE-based regret.} \label{fig:MRCP_NE}
\end{figure}

Despite the computation of MRCP in large games is infeasible due to the difficulty of repeatedly inquiring regret function, in matrix games, the regret inquiry is feasible.
To support the claim about the desired properties of MRCP for evaluation, in Fig.~\ref{fig:MRCP_NE}, we present results of averaged PSRO runs with FP and DO evaluated by MRCP-based regret based on a synthetic matrix game. 
We observe that the MRCP-based regret by definition is lower than its NE-based regret counterpart. Moreover, the NE-based regret reports the relative performance of theoretical ground truth. 
Notice that the gap between NE-based regret and MRCP-based regret dwindles as DO and FP gradually converge to true game NE, i.e., all regrets approach zero. 
We also observe that the MRCP-based regret curves are much smoother than the NE-based regret curves and indicates a steady performance improvement on the population of strategies. 

\subsection{MRCP Calculation Refinement}

In matrix games, MRCP can be computed by solving an optimization problem, i.e., follow the definition of MRCP, using amoeba method.
When applying the amoeba method to the MRCP optimization problem, we have to reconcile the fact that the optimization problem is constrained while the amoeba method is an unconstrained optimization technique. 
To solve this issue, Jordan \emph{et al.} \cite{Jordan10sw} propose a binary search to select the maximum feasible reflection and expansion scaling parameters (step sizes), respectively. 
However, this approach handles infeasibility by compromising the quality of the reflected and expanded points since the optimal solution points, which are high dimensional vectors, may not be reached given fixed scaling parameters. 
Motivated by the projection in constrained optimization, we fix this problem by projecting an infeasible point onto the unit simplex to maintain the feasibility. 
We show the performance of our algorithm dominates the previous one in Appendix~\ref{app: MRCP calculation}. 
Moreover, in Appendix~\ref{app: MRCP approx}, we investigate the possibility of approximating MRCP in games where the access to the regret function is limited and proposed an approximation approach.

\section{Evaluation without Exact Best Response}

As noted above, calculating profile regret for purposes of evaluating MSSs generally requires identifying a best-response strategy.
However, computing the exact best response may not be feasible in complex games. 
A particular approach is to collect the strategies generated across a set of PSRO runs, and evaluate regret with respect to that set.
This idea that evaluation is conducted based on generated strategies is general and has been widely applied especially when the true game is extremely large, e.g., the attack-graph game \citep{Wright19}. 
We refer to the game with all generated strategies as the \textit{combined game}. In general, regret with respect to the combined game is a lower bound on regret with respect to the true full game.
Since the combined game has been used in practice as a heuristic approach to evaluate strategy exploration, it is important to examine its effectiveness.

To test the effectiveness of this approach, we compare results for evaluation with respect to a combined game with that of exact best response (i.e., the ground truth in our context), for some games where calculating exact best responses is feasible. 
Results are shown in Fig.~\ref{fig:Combined Game regrets}.
We observe that high-regret profiles in the true game may exhibit quite low regret in the combined game. 
Most concerning is that the slack in the regret bound may vary across MSSs being evaluated, thus producing misleading comparisons.
Specifically in Fig.~\ref{fig:Combined Game regrets}, despite the apparent higher regret of FP profiles in the true game, FP profiles exhibit lower regret in the combined game. 
Our explanation for the phenomenon is that when one MSS can explore certain strategy to which strategies generated by other MSSs can deviate largely but not vice versa, the combined game fails to identify the correct ordering of MSSs.
Details of our combined-game analysis are provided in Appendix~\ref{app: combined game}.

\section{Conclusion}
The primary contributions of this study are methodological considerations for evaluating strategy exploration in EGTA, within the PSRO framework.
Our observations address nuances that have not been observed before, and may have led to misleading conclusions about the effectiveness of proposed methods. 
In particular, we propose an evaluation scheme with a consistency condition, dictating that progress in strategy exploration under different MSSs be evaluated with respect to the same solver. 
This condition, while seemingly obvious, has not always been followed, perhaps because it is natural in online learning settings to evaluate a method at any point based on its own solution criterion. 
In the context of strategy exploration, in contrast, what is important is not what the latest strategy is, but how it affects the solution of the model it is being added to.

We further investigate the significance of MRCP for strategy exploration, presenting evidence that the MRCP provides a principled basis for robust evaluation. We propose an algorithm to calculate MRCP that outperforms previous methods. 

Finally, we examine the effectiveness of the combined game for evaluating strategy exploration when calculating the exact best response is computationally infeasible. Based on our results, we cast doubt on the accuracy of this evaluation approach. 

\clearpage
\bibliographystyle{splncs04}
\bibliography{thebib, wellman}

\appendix
\clearpage
\newpage

\begin{center}
    \textbf{\Large{Appendices}}    
\end{center}

\section{Accuracy Improvement on MRCP Calculation}\label{app: MRCP calculation}

We show the procedure of projected amoeba method in Algorithm~\ref{alg:Projected Amoeba Method}.
Our goal is to minimize the cumulative regret function $f(\sigma)=\sum_{i\in N} \rho^{\mathcal{G}}_i(\sigma)$ shown in Equation~\ref{eq: MRCP def}. 
Denote the projection operator as $P(\sigma_i) = \argmin_{\sigma_i'\in \Delta(S_i)}{||\sigma_i' - \sigma_i||}$ for player $i\in N$. Denote $P(\sigma)$ as the projection operator for each $\sigma_i\in\sigma$. For amoeba method, we follow the default values of $\alpha =1$, $\gamma = 2$, $\rho =1/2$, $\sigma =1/2$.

\begin{algorithm}[!ht] 
\caption{Projected Amoeba Method}
\label{alg:Projected Amoeba Method}
\textbf{Input}: A full game model with regret function $f$ and an empirical game model.\\
\textbf{Parameter}: Amoeba method parameters $\alpha$, $\gamma$ , $\rho$  and $\sigma$ corresponding to the reflection, expansion, contraction and shrink coefficients.\\
\textbf{Output}: MRCP $\sigma$.
\begin{algorithmic}[1] 
\WHILE{t=1,...,T}
\STATE Select current test profiles $\sigma^1,...,\sigma^{n+1}$.
\STATE Order according to the regrets at these profiles: $f(\sigma^1)\le ... \le f(\sigma^{n+1})$.
\STATE Calculate $\sigma^o$, the centroid of profiles except $\sigma^{n+1}$
\STATE \textbf{Reflection:} Compute reflected point $\sigma^r = \sigma^o + \alpha(\sigma^o - \sigma^{n+1})$
\STATE Project $\sigma^r$ to probability simplex $\sigma^r = P(\sigma^r)$
\IF {$f(\sigma^1) \le f(\sigma^r) < f(\sigma^n)$}
\STATE $\sigma^{n+1} = \sigma^r$
\ELSE 
\STATE Continue.
\ENDIF
\STATE \textbf{Expansion:} 
\IF {$f(\sigma^r) < f(\sigma^1)$}
\STATE $\sigma^e = \sigma^o + \alpha(\sigma^r - \sigma^{o})$
\STATE $\sigma^e = P(\sigma^e)$
\IF {$f(\sigma^e) < f(\sigma^r)$}
\STATE $\sigma^{n+1} = \sigma^e$ and Continue.
\ELSE
\STATE $\sigma^{n+1} = \sigma^r$ and Continue.
\ENDIF
\ENDIF
\STATE \textbf{Contraction:} $\sigma^c = \sigma^o + \alpha(\sigma^{n+1} - \sigma^{o})$
\STATE $\sigma^c = P(\sigma^c)$
\IF {$f(\sigma^e) < f(\sigma^r)$}
\STATE $\sigma^{n+1} = \sigma^c$ and Continue.
\ENDIF
\STATE \textbf{Shrink:} $\sigma^c = \sigma^o + \alpha(\sigma^{n+1} - \sigma^{o})$
\STATE $\sigma^c = P(\sigma^c)$ and Continue.
\ENDWHILE
\STATE \textbf{return} $\sigma$
\end{algorithmic}
\end{algorithm}

We compare the performance of amoeba method with two approaches, i.e., BS and projection, in 2-player Kuhn's poker. 
Empirical games with different sizes are first sampled from the true game and then MRCP is approximated with different approaches. 
Table~\ref{tab:MRCP calculation comparison} shows the regret of MRCP calculated by different approaches. To illustrate the regret gap between different MRCP calculations, we also provide the regret of NE of the empirical game as a benchmark. 
We observe that for each size of a empirical game, calculating MRCP with projection results in a profile with significantly lower regret, merely with a different infeasibility handling approach. 
We also notice that when the size is small, the performance of two approaches is close. 
However, as the size increases, the BS approach does not lead to a good MRCP approximation and even NE of the empirical game could have lower regret than the MRCP approximation. 

Moreover, to understand the stability of results given by different approaches, we calculate the regret of approximated MRCP of a fixed empirical game for multiple times. 
We find that the variance of regrets given by BS approach from multiple runs is very large while the variance of our method is tiny. 
The improvement in the accuracy of MRCP approximation could dramatically benefit the study of evaluating strategy exploration since MRCP serves as a theoretical evaluation metric.

\begin{table*}[!hpbt]
  \centering
  \begin{tabular}{l ccccc| ccccc| ccccc}
    \toprule
     & \multicolumn{5}{c}{\textbf{Size = 5}} & \multicolumn{5}{c}{\textbf{Size = 7}} & \multicolumn{5}{c}{\textbf{Size = 9}}  \\
    \midrule
    \textbf{Index} & 1 & 2 & 3 & 4 & 5 & 1 & 2 & 3 & 4 & 5 & 1 & 2 & 3 & 4 & 5 \\
    \midrule
    \textbf{$\rho(\bar{\sigma})$ w. BS} & \textbf{0.39} & 0.36 & 0.35 & 0.44 & 0.19 & 0.51 & 0.36 & 0.44 & 0.39 & 0.21 & 0.26 & 0.40 & 0.44 & 0.45 & 0.83\\
    \textbf{$\rho(\bar{\sigma})$ w. Proj} & \textbf{0.39} & \textbf{0.30} & \textbf{0.30} & \textbf{0.40} & \textbf{0.19} & \textbf{0.31} & \textbf{0.30} & \textbf{0.32} & \textbf{0.35} & \textbf{0.14} & \textbf{0.15} & \textbf{0.33} & \textbf{0.33} & \textbf{0.38} & \textbf{0.61}\\
            \textbf{$\rho(\sigma^*)$} & 0.50 & 0.39 & 0.78 & 0.73 & 0.49 & 0.78 & 0.50 & 0.33 & 0.58 & 0.39 & 0.21 & \textbf{0.33} & 0.42 & 0.78 & 0.71\\
    \bottomrule
    \\
    \toprule
     & \multicolumn{5}{c}{\textbf{Size = 11}} & \multicolumn{5}{c}{\textbf{Size = 13}} & \multicolumn{5}{c}{\textbf{Size = 15}}  \\
    \midrule
    \textbf{Index} & 1 & 2 & 3 & 4 & 5 & 1 & 2 & 3 & 4 & 5 & 1 & 2 & 3 & 4 & 5 \\
    \midrule
    \textbf{$\rho(\bar{\sigma})$ w. BS} & 0.46 & 0.49 & 0.45 & 0.59 & 0.60 & 0.37 & 0.57 & 0.60 & 0.28 & 0.27 & 0.37 & 0.52 & 0.30 & 0.27 & 0.17\\
    \textbf{$\rho(\bar{\sigma})$ w. Proj} & \textbf{0.07} & \textbf{0.35} & \textbf{0.17} & \textbf{0.30} & \textbf{0.35} & \textbf{0.13} & \textbf{0.33} & \textbf{0.28} & \textbf{0.12} & \textbf{0.18} & \textbf{0.08} & \textbf{0.17} & \textbf{0.08} & \textbf{0.16} & \textbf{0.07}\\
    \textbf{$\rho(\sigma^*)$} & 0.29 & 0.50 & 0.26 & 0.33 & 0.67 & 0.22 & 0.53 & 0.50 & 0.20 & 0.28 & 0.50 & 0.57 & 0.20 & 0.27 & 0.19\\
    \bottomrule
  \end{tabular}
  \caption{MRCP quality with different infeasibility handling methods.}
  \label{tab:MRCP calculation comparison}
\end{table*}

\section{MRCP Approximation} \label{app: MRCP approx}

Calculating MRCP in large games can be infeasible since it requires repeatedly regret calculations, and thus expensive best response operations, which is completely not affordable. 
To approximate the true MRCP of the empirical game game, we propose an algorithm that estimates the regret of every mixed strategy profile needed for MRCP search, which is a key step for MRCP calculation.
Specifically, we derive an upper bound for the regret of a mixed strategy profile through the deviation payoff of a finite set of pure strategy profiles in the empirical game. 
Therefore, we reduce the regret calculation of infinite number of mixed strategy profiles to that of finite number of pure strategy profiles.
Mathematically, we have
\begin{align}
    \rho^{\mathcal{G}}_i(\sigma) &= \max_{s_i'\in S_i}u_i(s_i', \sigma_{-i})-u_i(\sigma_i, \sigma_{-i})\\
    & = \max_{s_i'\in S_i}\sum_{s_{-i}\in S_{-i}}\sigma(s_{-i})u_i(s_i', s_{-i})\\
    & -\sum_{s_i\in S_i}\sum_{s_{-i}\in S_{-i}}\sigma(s_i)\sigma(s_{-i})u_i(s_i, s_{-i})\\
    & \le \sum_{s_{-i}\in S_{-i}}\sigma(s_{-i})\max_{s_i'\in S_i} u_i(s_i', s_{-i})\\
    & -\sum_{s_i\in S_i}\sum_{s_{-i}\in S_{-i}}\sigma(s_i)\sigma(s_{-i})u_i(s_i, s_{-i}) \label{eq:regret in bound}
\end{align}
Note that to calculate the deviation payoff $\max_{s_i'\in S_i} u_i(s_i', s_{-i})$, there is no need to best respond to all pure strategy profiles, whose number could be very large as the empirical game is being extended, but only the other player's strategies since given other players' strategies fixed, the deviation payoffs of all strategies of current player are equal. 
This approximation avoids repeated calling of best response oracle for the calculation of regret of mixed strategy profiles. 
In other words, we only need $\sum_{i\in N}|S_i|$ calculations of regret compared to thousands of calculations without approximation. 

However, this approximation may suffer issues in some games, which could lead to a failure. 
First, the utility structure of a game affects the performance of our approximation method. 
For example, consider a zero-sum game in which the utilities of each strategy profile are generated according to a uniform distribution with upper bound $R$. 
When the strategy space of the game is large, it is very likely that 
\begin{displaymath}
     \max_{s_i'\in S_i} u_i(s_i', s_{-i}) \approx R, \forall s_{-i}\in S_{-i}
\end{displaymath}
Then solving equation~\ref{eq:regret in bound} is equivalent to find the strategy profile that maximizes the sum of players' utilities. 
However, since the game is zero-sum, the sum of players' utilities is zero for every profile and the term in~\ref{eq:regret in bound} is canceled. 
So the upper bound of regret of any profile is equal to $R$, which would fail the approximation. 
Moreover, since the sum of players' utilities is zero for every profile in zero-sum game, the profile given by equation~\ref{eq:MRCP} is always a pure strategy profile and this could result in a large estimation error.

We handle the potential failure for zero-sum games by replacing the sum of regret over players in equation~\ref{eq:MRCP} with the maximal regret over players. 
Mathematically, we optimize the following for MRCP $\Tilde{\sigma}$. 
\begin{equation}\label{eq:MRCP}
\Tilde{\sigma}^{X} = \argmin_{\sigma\in \Delta(X)} \max_{i\in N} \rho^{\mathcal{G}}_i(\sigma)
\end{equation}
In other words, we minimize the maximal regret over players to approximate the true MRCP $\bar{\sigma}$. This modification prevents the term~\ref{eq:regret in bound}, which serves to make the output profile away from pure strategy profile, from being offset.

To verify that the two definitions of MRCP have the similar quality, i.e. the regrets of these two MRCPs are close, we calculate the two MRCPs in Kuhn's poker and analyze their performance. 
Note that the regret of MRCP $\bar{\sigma}$ is the lower bound of $\Tilde{\sigma}$.

To solve for MRCP, we consider a normal-form representation of Kuhn's poker and randomly sample empirical games with different sizes, i.e. the number of strategies per player, from the true game. 
For each size, we sample 5 different empirical games. 
We calculate the MRCP with two definitions for each empirical game and compare the sum of regret over players of the output profiles. 
To illustrate the regret gap between two definitions is sufficiently small, we also provide the regret of NE of the empirical game as a benchmark. 
Table~\ref{tab:MRCP quality} shows that for each size of a empirical game, calculating MRCP with two definitions results in profiles with similar quality.

\begin{table*}[!tbpt]
  \centering
  \begin{tabular}{l ccccc| ccccc| ccccc}
    \toprule
     & \multicolumn{5}{c}{\textbf{Size = 5}} & \multicolumn{5}{c}{\textbf{Size = 10}} & \multicolumn{5}{c}{\textbf{Size = 15}}  \\
    \midrule
    \textbf{Index} & 1 & 2 & 3 & 4 & 5 & 1 & 2 & 3 & 4 & 5 & 1 & 2 & 3 & 4 & 5 \\
    \midrule
     $\rho(\bar{\sigma}$) & 0.67 & 0.25  & 0.58 & 0.20 & 0.58 & 0.09  & 0.20  & 0.15 & 0.33  & 0.33 & 0.05& 0.08 & 0.07& 0.38 & 0.17\\
    $\rho(\Tilde{\sigma})$ & 0.67 & 0.27 & 0.63 & 0.24 & 0.65 & 0.09 & 0.20 & 0.15 & 0.34  & 0.36 & 0.07 & 0.09 & 0.09& 0.40 & 0.21\\
    $\rho(\sigma^*) $ & 0.83 & 0.50 & 0.72 & 0.42 & 0.83 & 0.17 & 0.44 & 0.26 & 0.54 & 0.61 & 0.26 & 0.21 & 0.16 & 0.46 & 0.46\\
    \bottomrule
  \end{tabular}
  \caption{MRCP quality with two definitions.}
  \label{tab:MRCP quality}
\end{table*}

\subsection{Experimental Results}
\begin{table*}[!tbpt]
  \centering
  \begin{tabular}{l ccccc| ccccc| ccccc}
    \toprule
     & \multicolumn{5}{c}{\textbf{Size = 3}} & \multicolumn{5}{c}{\textbf{Size = 5}} & \multicolumn{5}{c}{\textbf{Size = 7}}  \\
    \midrule
    \textbf{Index} & 1 & 2 & 3 & 4 & 5 & 1 & 2 & 3 & 4 & 5 & 1 & 2 & 3 & 4 & 5 \\
    \midrule
      $\rho(\bar{\sigma}$) & 359 & 262 & 232 & 428 & 305 & 176 & 124 & 487 & 364 & 75 & 95 & 228 & 627 & 103 & 322\\
    $\rho(\Tilde{\sigma})$ & 505 & 275 & 265 & 532 & 353 & 253 & 144 & 727 & 365 & 106 & 575 & 397 & 794 & 183 & 322\\
         $\rho(\sigma^*) $ & 615 & 275 & 242 & 554 & 949 & 535 & 144 & 806 & 737 & 377 & 491 & 514 & 973 & 172 & 507\\
    \bottomrule
    \\
    \toprule
     & \multicolumn{5}{c}{\textbf{Size = 9}} & \multicolumn{5}{c}{\textbf{Size = 11}} & \multicolumn{5}{c}{\textbf{Size = 13}}  \\
    \midrule
    \textbf{Index} & 1 & 2 & 3 & 4 & 5 & 1 & 2 & 3 & 4 & 5 & 1 & 2 & 3 & 4 & 5 \\
    \midrule
      $\rho(\bar{\sigma}$) & 160 & 121 & 180 & 181 & 17 & 247 & 250 & 243 & 68 & 108 & 324 & 60 & 209 & 103 & 204\\
    $\rho(\Tilde{\sigma})$ & 249 & 156 & 205 & 230 & 21 & 263 & 405 & 378 & 165 & 108 & 435 & 60 & 318 & 134 & 483\\
         $\rho(\sigma^*) $ & 236 & 314 & 759 & 330 & 420 & 388 & 596 & 446 & 152 & 646 & 705 & 216 & 327 & 479 & 204\\
    \bottomrule
    
  \end{tabular}
  \caption{MRCP quality with approximation in symmetric-zero sum game.}
  \label{tab:MRCP approximation}
\end{table*}

With the new definition of MRCP, we measure the quality of approximated MRCP in a synthetic two-player zero-sum game in which the regrets of MRCP $\bar{\sigma}$, approximated MRCP $\Tilde{\sigma}$ (we reload the notation for convenience) and NE $\sigma^*$ are compared, shown in Table~\ref{tab:MRCP approximation}. We find that in some sampled empirical games, the approximation gives profiles with very similar regret as that of the MRCP\@.
However, we observe that the good approximation is not consistent across all sampled empirical games. 
One explanation is that when the true MRCP is close to a pure strategy profile, the approximation gives good results while the performance of approximation declines when the randomness in MRCP is high. 
Future study on this thread could include tightening the bound to get better approximation of MRCP.

\section{The Combined Game Example} \label{app: combined game}

The exact best response oracle provides a regret calculation tool for evaluating MSSs. Despite its availability in the poker games that we experiment with, in many complex games, calculation for the exact best response is not feasible due to the large game trees. 
An alternative that have been employed in many prior works is applying the concept of the combined game as a heuristic evaluation approach. 
In this section, we verify its effectiveness from a game-theoretic perspective.

A combined game is an encompassing empirical game whose strategy set is a union over all strategy sets of the empirical games that are generated across different MSSs or random seeds. The missing payoff entries are further simulated. 
Evaluation is conducted by viewing the combined game as the approximation of the true game and regret calculation only considers deviations within the combined game. The general idea of the combined game, i.e., using generated strategies for evaluation purpose, is widely applied and can be viewed as an instance of agent vs agent by Balduzzi \emph{et al.} \cite{balduzzi2018re}.

We test this approach in games where exact best responses are available and thus the ground truth performance of different MSSs is known. We consider three MSS PRD, NE and uniform in 2-player Leduc poker. For each MSS, we perform 3 differently seeded PSRO runs with 150 strategies, thus each player has 1350 strategies in the combined game. We compare results of evaluation conducted with the combined game and the exact best response. In Fig.~\ref{fig:Combined Game regrets}, each curve is an average of three runs with different seeds. The three averaged NE-based regret curves on the top are generated with the exact best response oracle, i.e., evaluating with respect to the true game, while the bottom ones are created by only considering deviations within the combined game. The stratification is caused by the fact that the regret of a profile in the true game is lower bounded by the empirical game.

We observe that the regret curves based on the combined game does not truthfully reflect the order of performance with exact best response. In Fig.~\ref{fig:Combined Game regrets}, despite the apparent higher regret of FP profiles in the true game, FP profiles exhibit low regret in the combined game. 
The disparity might stem from the strategy generation process of FP: new strategies are trained against a uniform opponent strategy, which makes it harder to be exploited by the counterpart of DO and PRD. 
Based on our results, we cast doubt on the accuracy of this evaluation approach. 

\begin{figure}[!ht]
\centering
\includegraphics[width=6cm]{ 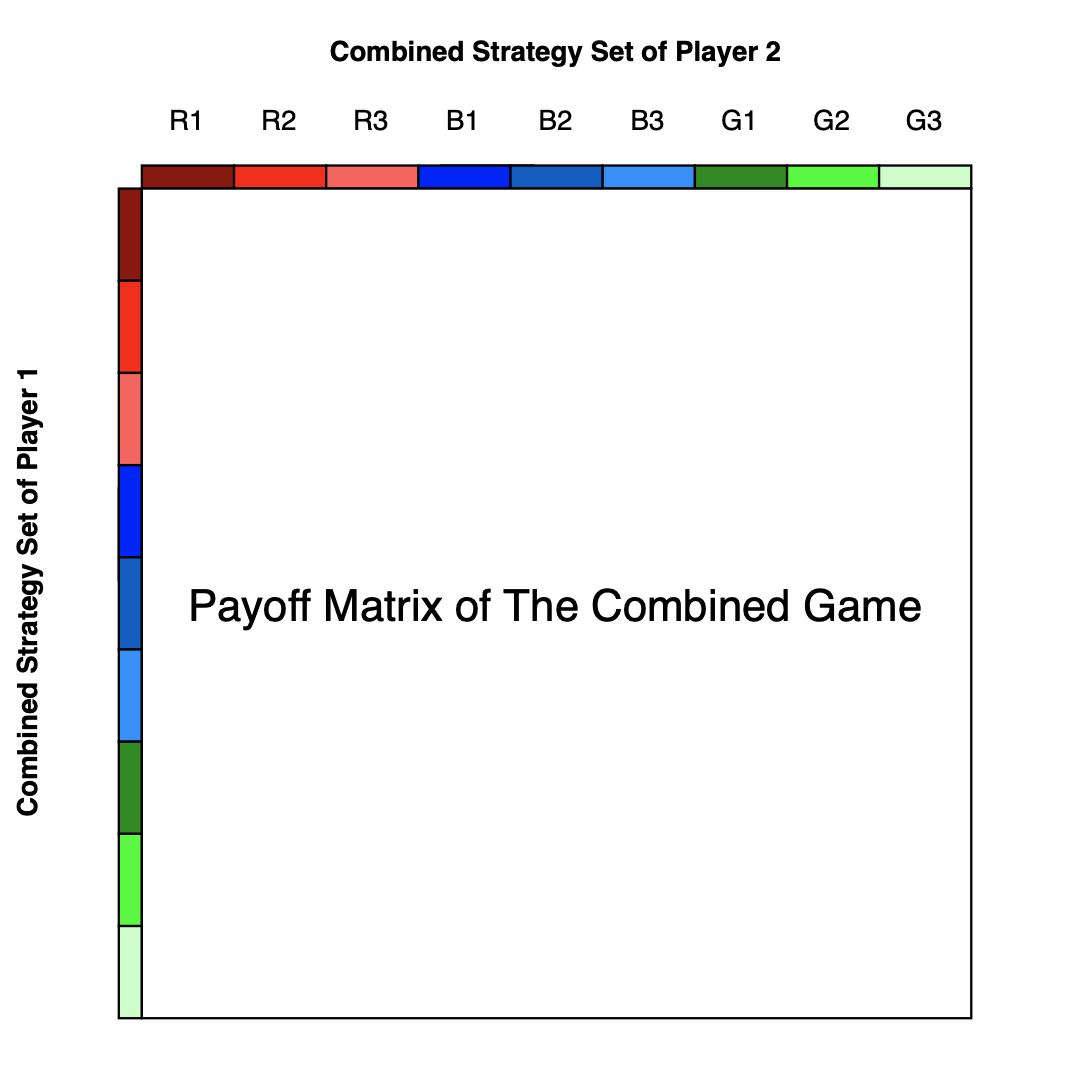}
\caption{Combined Game Example } \label{fig:combined_game}
\end{figure}

In fact, the combined game essentially attempts to utilize partial strategy space to approximate the relative performance in the global, which does not hold in general. 
But in certain types of game, for example, in transitive games, local relative performance does indicate the global performance.
Finally we give an illustration of the idea of combined game. Suppose we run three MSSs (R, B, G) in a 2-player game. For each MSS, we take 3 runs whose strategies are assigned different darkness. So the payoff matrix of the combined game is shown in Fig.~\ref{fig:combined_game}.

\section{Choice of Solution Concepts in Empirical Game} \label{app: solutions}
In EGTA, an empirical game is used to approximate the true game and NE of the empirical game is viewed as an approximate NE to the true game. 
In some scenarios, e.g., games with large strategy space, limited number of strategies are available to be considered in the empirical game due to constraints on computational resources or the difficulty of hand-crafting strategies. 
When the true game may not be well-approximated, we pose a following question: is NE of the empirical game truly the best solution concept to adopt?

Through our experiments, we notice that the empirical NE could have quite large regret in the true game compared with other profiles in the empirical game. 
Even worse, it is possible that the empirical NE possesses the largest regret among all possible profiles in some extreme cases. 
Consider matrix game Table~\ref{tab: Matrix game for unstable NE} with empirical game $(a^2_1, a^2_2)$ for both players. The NE $(a^2_1, a^2_2)$ is the most unstable pure strategy profile in the empirical game. Table~\ref{tab: regrets} list the regret of each pure strategy profile.

\begin{table}[!ht]
    \centering
    \begin{tabular}{ |c|c|c|c| } 
 \hline
  & $a_2^1$ & $a_2^2$ & $a_2^3$  \\ 
 \hline
 $a_1^1$ & (0, 0) & (-1, 1) & (-2, 2) \\ 
 \hline
 $a_1^2$ & (1, -1) & (0, 0) & (-5, 5) \\ 
 \hline
 $a_1^3$ & (2, -2) & (5, -5) & (0, 0) \\ 
 \hline
\end{tabular}
\caption{Handcrafted Symmetric Zero-Sum Game.}
\label{tab: Matrix game for unstable NE}
\end{table}{}

\begin{table}[!ht]
    \centering
    \begin{tabular}{ cc} 
 \toprule
 \textbf{Profiles} &  \textbf{regret}   \\ 
 \midrule
 $(a^1_1, a^1_2)$ & 4  \\ 
 $(a^2_1, a^1_2)$ & 7 \\
 $(a^1_1, a^2_2)$ & 7  \\
 $(a^2_1, a^2_2)$ & 10 \\
 \bottomrule
\end{tabular}
\caption{Regrets of Pure Strategy Profiles.}
\label{tab: regrets}
\end{table}{}

The example inspires us to reconsider the solution concept to apply to real world games: NE strategies of the empirical game could be far from optimum in a global perspective and regularization to NE may be required to obtain a stable performance.

\section{Experimental Parameters}\label{app: hyperparams}
\subsection{Synthetic Games}
In the experiment shown in Fig.~\ref{fig:Difference between 2 NashConv Matrix Game}, we have 2-player zero-sum matrix games with 100 strategies per player with uniformly sampled integer payoffs from $-10$ to $10$, both inclusive. 

In the experiment shown in Fig.~\ref{fig:MRCP_NE}, we have 2-player zero-sum matrix games with 200 strategies per player with uniformly sampled integer payoffs from $-10$ to $10$, both inclusive. 

In the MRCP approximation experiment, we have 2-player zero-sum matrix games with 200 strategies per player with uniformly sampled integer payoffs from $-1000$ to $1000$, both inclusive. The range of payoffs is enlarged to prevent same payoffs assigned to large number of actions.

\subsection{Poker Games}
We use OpenSpiel \citep{lanctot2019openspiel} default parameter sets for experiments on Leduc and Kuhn's poker: each payoff entry in an empirical game is an average of 1000 repeated simulations; DQN is adopted as a best response oracle, its parameters are shown in Table~\ref{tab:dqn_parameter}. The poker games are asymmetric in the sense that one player always moves first.

PRD is implemented with lower bound for strategy probability 1e-10, maximum number of steps 1e5 and step size 1e-3. 

\begin{table}[!htb]
    \centering
    \begin{tabular}{lc}
        \toprule
        \textbf{Parameter} & \textbf{Value}\\
        \midrule
        learning rate & 1e-2\\
        Batch Size & 32\\
        Replay Buffer Size & 1e4\\
        Episodes & 1e4\\
        optimizer & adam\\
        layer size & 256\\
        number of layer & 4\\
        Epsilon Start & 1\\
        Epsilon End & 0.1\\
        Exploration Decay Duration & 3e6\\
        discount factor & 0.999\\
        network update step & 10\\
        target network update steps & 500\\
        \bottomrule
    \end{tabular}
    \caption{DQN parameter}
    \label{tab:dqn_parameter}
\end{table}

\end{document}